\documentclass[aps,prl,twocolumn,amsfonts,showpacs]{revtex4}
\usepackage{epsfig}
\usepackage{amssymb,amsmath,amsfonts,hyperref}
\usepackage{wasysym}
\usepackage{latexsym}
\newcommand{\gae}{\lower 2pt \hbox{$\, \buildrel {\scriptstyle >}\over {\scriptstyle
\sim}\,$}}
\newcommand{\lae}{\lower 2pt \hbox{$\, \buildrel {\scriptstyle <}\over {\scriptstyle
\sim}\,$}}
\begin{document}
\title{Finite-size effects on the dynamics of the zero-range process}
\author { Shamik Gupta,$^1$  Mustansir Barma,$^1$ and Satya N. Majumdar$^2$}
\address{
{\small $^1$Department of Theoretical Physics, Tata Institute of
Fundamental Research, Homi Bhabha Road, Mumbai, India}\\
{\small $^2$Laboratoire de Physique Th\'eorique et Mod\`eles Statistiques,
        Universit\'e Paris-Sud, Orsay, France}}
\date{\today}
\begin{abstract}
We study finite-size effects on the dynamics of a one-dimensional
zero-range process which shows a phase transition from a low-density
disordered phase to a high-density condensed phase. The
current fluctuations in the steady state show
striking differences in the two phases.  In the
disordered phase, the variance of the integrated current shows damped
oscillations in time due to the motion of fluctuations
around the ring as a dissipating kinematic wave. In the
condensed phase, this wave cannot propagate through the condensate, and
the dynamics is dominated by the long-time relocation of the
condensate from site to site. 
\end{abstract}
\pacs{05.70.Ln, 02.50.Ey, 05.40.-a, 64.60.-i}
\maketitle

The dynamics of fluctuations in simple nonequilibrium steady states of
interacting particle systems has
been studied extensively in recent years, and a fairly good
understanding of the physical processes involved has been achieved in infinite systems
\cite{review}.  
Studies of these fluctuations in a finite system show a strong imprint of the nonequilibrium
character, which combines with size effects to bring in interesting
dynamical phenomena \cite{sg}. Many of these systems often exhibit
a nonequilibrium phase transition in the steady state, from a disordered
to an ordered phase, as one tunes
an external parameter such as the density. 
This raises the natural question: How is the dynamics of fluctuations in
the steady state affected 
as the system passes through a
nonequilibrium phase transition?
In this paper, we address this
issue within the ambit of a paradigmatic model, the zero-range process (ZRP)
\cite{spitzer,zrprev}, and show that there are strong differences in
the dynamical properties arising from very different physical processes
in the two phases. 

The ZRP involves biased hopping of particles on a periodic lattice with
a rate that depends on the occupancy at the
departure site. At long times, the system reaches a
nonequilibrium steady state.
For certain classes of the hop rates, as the particle density crosses a critical threshold, 
there is a continuous phase transition from a disordered phase
with uniform average density to a condensed phase where a finite
fraction of particles (the
``condensate") accumulates on a single site \cite{zrprev}. 
The ZRP applies to a wide variety of physical systems ranging
from traffic flow \cite{traffic} to shaken granular gases
\cite{granular}; in addition, it was invoked to provide a 
criterion for phase separation in one-dimensional driven systems \cite{phasesep}.

\begin{figure}[h!]
\begin{center}
\includegraphics[scale=0.35]{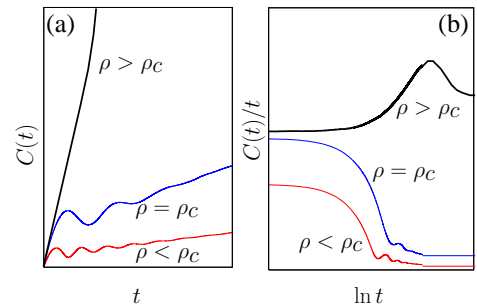}
\caption{ (Color online) (a) Schematic plot of the integrated current fluctuations $C(t)$ across a
bond in the steady state of the ZRP as a function of time $t$ in the
disordered phase ($\rho < \rho_{c}$), at the critical point
($\rho=\rho_{c}$), and in the condensed phase ($\rho > \rho_{c}$). (b)
Schematic plot of $C(t)/t$ as a function of $t$ in all the phases.}
\label{fig1} 
\end{center}
\end{figure}

Recent work on the ZRP has dealt with the relaxation of
an initial homogeneous density distribution toward the condensed phase
\cite{dynnonst}. By contrast, here we are interested in
the dynamics of density fluctuations in the {\it steady state} in 
both the disordered and the condensed phases. We explore the dynamics by monitoring the 
variance of the integrated particle current in Monte Carlo simulations
and supplement our findings by analyzing the survival probability
distribution of the largest mass in the system. The relevant time scales
in the behavior of fluctuations in the two phases and the physical effects
underlying them are summarized below.

In Fig. \ref{fig1}, we show schematically the behavior of the variance
of the integrated current as a function of time in different phases. In
all the phases, at short times, the integrated current is
Poisson distributed, implying that the variance grows linearly in time,
a behavior which holds for all times in an infinite system. In a finite
system, in the disordered phase and at criticality, the variance shows
oscillations at times proportional to the system size $L$. This results from kinematic waves transporting density fluctuations around the system with a well-defined
speed. At longer times ($\sim L^{3/2}$), the wave decays, and
then the variance increases linearly with time with a small slope that decreases with
increasing system size. In the condensed phase, however, the kinematic
wave cannot pass through the condensate; thus, fluctuations do not
circulate around. The initial linear behavior continues until, after a
characteristic time which grows as a power of the system size, the
condensate relocates itself. This results in the variance showing a 
linear rise in time with a much larger slope than at early times.
Subsequently, after the condensate has relocated to another site, the
slope of the linear rise slowly approaches a size-dependent constant.  

We now turn to a derivation of these properties. The ZRP involves $N$ particles of 
unit mass on a ring of size $L$ with arbitrary
 occupancy allowed at any site. A particle hops out of a randomly
 selected site $i$ with occupancy $n_{i}$ with a specified rate
 $u(n_{i})$, and goes to site $(i+1)$. In the thermodynamic
 limit $N \rightarrow \infty, L \rightarrow \infty$ at a fixed density
 $\rho=N/L$, the probability $P(\mathbf{C})$ of a configuration
 $\mathbf{C} \equiv
 \{n_{i}\}$ in the steady state, in the grand canonical ensemble, is given by \cite{zrprev} 
\begin{equation}
P(\mathbf{C})=\frac{1}{Z}\prod_{i=1}^{L}v^{n_{i}}f(n_{i}), \nonumber \\
\end{equation}
\begin{equation}
f(n)=\left\{ 
\begin{array}{ll}
               \left( \prod\limits_{l=1}^{n}u(l) \right)^{-1} & \mbox{if $n>0$}, \\
               1 & \mbox{if $n=0$}.
               \end{array}
        \right. \\
\label{steadystate}
\end{equation}
Here, $Z$ is the partition function and $v$ is the fugacity. The
average mass at a site equals $vF'(v)/F(v)$, where
$F(v)=1+\sum_{n=1}^{\infty}v^{n}f(n)$. Particle conservation gives a
relation between $\rho$ and $v$, namely, $\rho=vF'(v)/F(v)$. The fugacity has the maximum
value $v_{\mathrm{max}}=u(\infty)$, given by the radius of convergence of the infinite series
$F(v)$. The ZRP can be mapped to a generalization of the asymmetric simple exclusion process (ASEP) by interpreting the ZRP sites as particles in the ASEP,
while the particles at a ZRP site become holes preceding the
corresponding ASEP particle \cite{zrprev}. The hop rate $u(n)$ for an
ASEP particle, now a function of the headway to the next particle,
induces a long-ranged particle hopping.  

Here, we consider the hop rate $u(n)=1+b/n$ with $b>2$
for which the system undergoes a nonequilibrium phase transition
\cite{zrprev}. As $\rho$ crosses the critical value $\rho_{c}=1/(b-2)$
\cite{dynnonst}, a low-density
disordered phase with mass of $O(1)$ at each site evolves to a
high-density condensed phase where a macroscopic collection of particles of average
mass $(\rho-\rho_{c})L$ condenses onto a randomly selected site, while the
remaining sites have the average mass $\rho_{c}$. 

In the steady state, the mean current of particles between every pair of
neighboring sites is
the same. In the thermodynamic limit, the mean current is 
$J=\sum_{n=1}^{\infty}u(n)v^{n}f(n)/F(v)=v$. It increases with the
density $\rho$, attaining its maximum value  $v_{\mathrm{max}}=u(\infty)=1$ at $\rho_{c}$, and remaining pinned to this value in
the condensed phase. 
To address the dynamics of density 
fluctuations, we examine the fluctuations in the integrated particle current across any bond in the
steady state. The large deviation function of the integrated current has been studied for the ZRP with open boundaries in \cite{zrpopld}. For our
purpose, we monitor the variance $C(i,t)$ of the integrated current
$H(i,t)$ which counts the total number of particles crossing
the bond $(i, i+1)$ in time $t$. Thus, we have  
\begin{equation}
C(i,t)\equiv\langle H^{2}(i,t) \rangle - \langle H(i,t) \rangle^{2}.
\end{equation}
In the equivalent ASEP, $C(i,t)$ is a measure of tagged
particle correlations, being given by the variance of
the $i$th tagged particle around its average displacement in time $t$. 
In simulations, we monitor $C(t)=\sum_{i=1}^{L}C(i,t)/L$ and find that
it shows strong differences in
behavior in the disordered and the condensed phases, reflecting 
very different underlying physical processes in the two phases.

\textit{Disordered phase.} $C(t)$ in this phase behaves similarly to the tagged particle correlation in the ordinary ASEP, studied recently \cite{sg}. As discussed below, there are two scales $T_{1} \sim L$, set
by the circulation time of a kinematic wave of density fluctuations, and
$T_{2} \sim L^{3/2}$, given by the time taken by this wave to decay.

(i) $ t \ll T_{1}$. Here, $C(t)$ grows linearly in time: $C(t)=vt$. This
follows from the result that, in this time regime, $H(i,t)$ is
Poisson distributed with intensity $v$ over all bonds $(i, i+1)$. 
The population at a ZRP site undergoes a time-reversible birth-death
process where a birth (particle input) occurs with rate $v$,
while a population of $n$ particles undergoes a death with rate
$u(n)$. For a reversible birth-death process with Poisson inputs, Burke's theorem
implies an identical Poisson distribution of outputs \cite{burke}. Noting
that the output from one site forms the input to the next
site then implies the result \cite{lebowitz}.

(ii) $ T_{1} \ll t \ll T_{2}$. In this regime, $C(t)$ oscillates as a
function of time. In a driven system with homogeneous density $\rho$ and a density-dependent current
$J(\rho)$, density fluctuations are transported as a kinematic wave with speed $v_{K}=\partial J/\partial \rho$
\cite{lighthill}. This wave is dissipated over a time scale $\sim
L^{z}$, where $z$ is the dynamic exponent of the system. For the ZRP,
$z$ takes on the Kardar-Parisi-Zhang (KPZ) value of $3/2$ \cite{kpz}.
Since $z >1$, fluctuations circulate several times around a periodic
system before getting dissipated, and revisit every site after a time $L/v_{K}$.
This makes the variance oscillate in time with this period. A
measure of the growth of dissipation in time is given by the lower
envelope of the oscillations, which behaves as $t^{2\beta}$ where
$\beta=\beta_{\mathrm{KPZ}}=1/3$ \cite{kpz}. 

(iii) $ t\gg T_{2}$. The time scale $T_{2}$ marks the dissipation time
of an initial density profile. For times $t
\gg T_{2}$, the variance grows diffusively: $C(t) \sim D(L)t$. Matching this behavior
 at $T_{2}$ with that in (ii) above gives $D(L) \sim L^{-1/2}$, as for
 the ordinary ASEP \cite{derrida}. 

\begin{figure}
\includegraphics[scale=0.35]{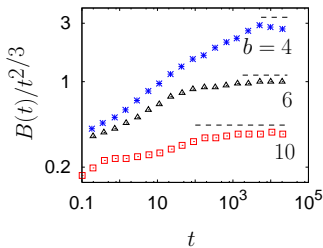}
\caption{ (Color online) Monte Carlo (MC) simulation results for
$B(t)/t^{2/3}$ as a function of time $t$ for different $b$'s at
$\rho_{c}=1/(b-2)$. The system size is $L=16 ~384$. The dashed lines show the asymptotic $t^{2/3}$ approach.}
\label{btcond}
\end{figure}

\textit{Critical point}. At the critical density $\rho_{c}$, the variance behaves differently 
for values of $b \le 3$ and $b > 3$. For $b \le
3$, there is no moving kinematic wave \cite{dynnonst}. Hence, the
integrated current is Poisson distributed with intensity
$v_{\mathrm{max}}=1$,
implying that the variance continues to grow linearly with slope $1$. For $b >3$, however, the kinematic wave
speed is nonzero and the Poisson distribution for the integrated
current is expected to hold for times smaller than the return time of
the kinematic wave. At criticality, the largest mass in the system
scales as $\sim L^{1/(b-1)}$  \cite{footnote1} and is insufficient to block the
circulation of the kinematic wave around the system. $C(t)$ oscillates
in time as for $\rho < \rho_{c}$,
with return time $T_{1}$ and decay time $T_{2}$ of the kinematic wave.
To find the exponent $\beta$ at criticality, we monitored the variance
$B(t)$ of the
integrated current by starting from an arbitrary but \textit{fixed} initial configuration, drawn from
the stationary ensemble \cite{vanb}. $B(t)$ grows asymptotically as
$t^{2\beta}$ for $t \ll L^{3/2}$ \cite{sg}. We find that $\beta$ at
criticality has the KPZ value of $1/3$, independent
of $b$ (Fig. \ref{btcond}).

\begin{figure}
\begin{center}
\includegraphics[scale=0.35]{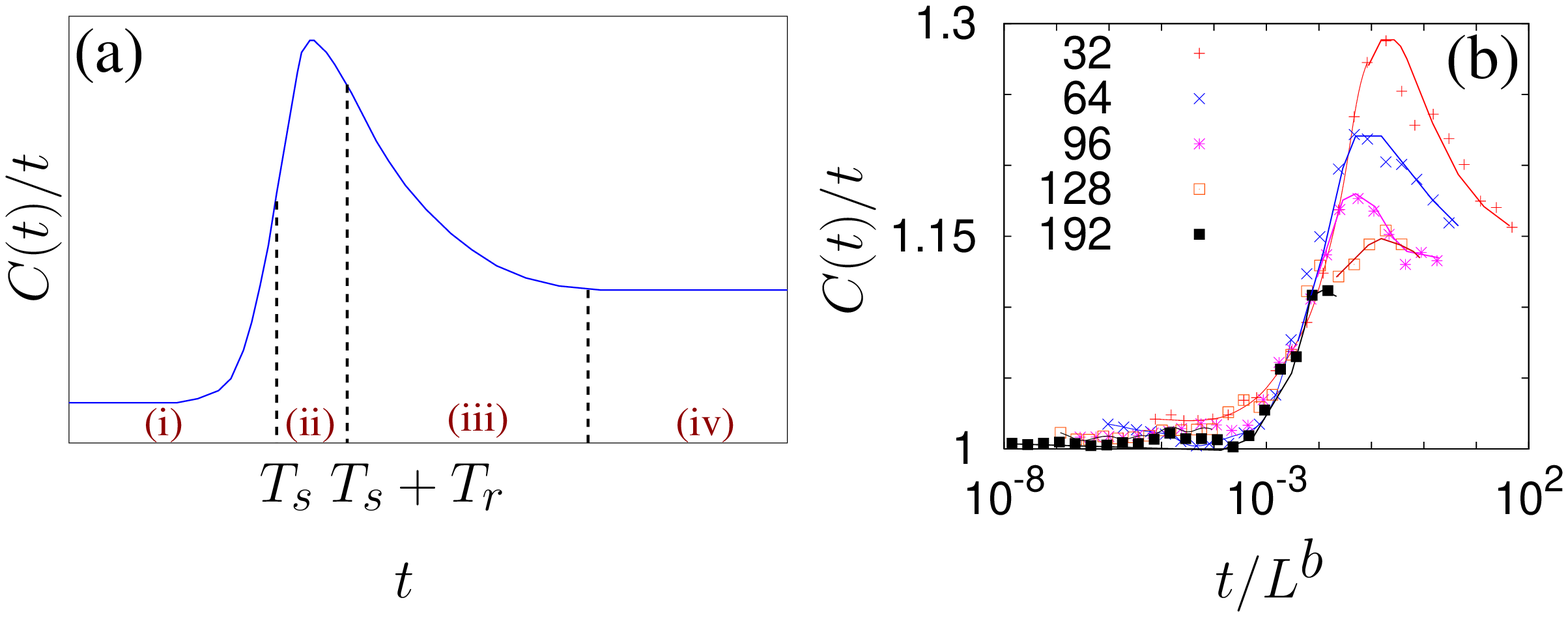}
\caption{ (Color online) (a) Schematic plot of $C(t)/t$ as a function of
time $t$ in the condensed phase. The various regimes (see
text) are also marked. (b) Scaling of $C(t)/t$ with $t/L^{b}$
for $b=3$ in the condensed phase ($\rho=4$). Various system sizes $L$
considered are marked in the figure. The data points, obtained
from MC simulations, are
connected by smooth curves.}
 \label{ctcond}
\end{center}
\end{figure}

\textit{Condensed phase}. For $\rho > \rho_{c}$, a finite fraction of
the total mass (the condensate) resides on one site for the characteristic survival time $T_{s} \sim
(\rho-\rho_{c})^{b+1}L^{b}$ \cite{godrechezrp}. It then relocates to
another site over the relocation time scale $T_{r} \sim
(\rho-\rho_{c})^{2}L^{2}$, as discussed below. The behavior of $C(t)$ in
this phase is best depicted by plotting
$C(t)/t$ as a function of time, as shown schematically in Fig.
\ref{ctcond}(a), where the various regimes are also marked. 

(i) $t \ll T_{s}$: Here, $C(t)/t$ equals $1$, with a mild upward deviation for longer times. (ii) $t \sim
T_{s}$: In this regime, $C(t)/t$ rises rapidly in time. (iii) $t
\gae T_{s}+T_{r}$: Here, $C(t)/t$ falls slowly in
time. (iv) $t \gg T_{s}+T_{r}$: Here, $C(t)/t$ begins to approach a
size-dependent constant, as predicted by a simple model described later
in this paper. Features (ii), (iii), and (iv) result from enhanced fluctuations due to the condensate relocation. To understand this, we need to first discuss the relocation dynamics.   

The condensate relocation occurs
through exchange of particles between two
sites. On monitoring the time evolution of the largest and the second largest mass in
simulation, the following picture emerges. Let $M(t)$ denote the
largest mass in the steady state at time $t$. $M(t)$ has
the average value $M_{0} \equiv \langle M(t) \rangle =
(\rho-\rho_{c})L$, with fluctuations $\Delta M_{0}$ which scale as $L^{1/2}$ for $b >
3$, and as $L^{1/(b-1)}$ for $2 < b <3$ \cite{satyacond}. These
fluctuations may build up in time, and
over the time scale $T_{s}$, the largest mass fluctuates to 
$\sim M_{0}/2$, while a mass $\sim M_{0}/2$ also builds up at another
site. Subsequent to this, two sites with mass $\sim M_{0}/2$ exchange particles between themselves
resulting in relatively rapid alternating relocations of the largest
mass from one site to the other. The difference of masses on these two sites
performs an unbiased random walk in time until fluctuations populate one
of the sites to $\sim M_{0}$ at the expense of the other, which happens
over the time scale $T_{r} \sim (\rho-\rho_{c})^{2}L^{2}$.
\begin{figure}
\begin{center}
\includegraphics[scale=0.35]{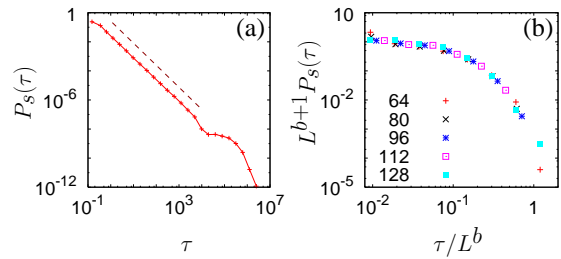}
\caption{(Color online) (a) Survival probability distribution $P_{s}(\tau)$ of the largest
mass in the condensed phase. The system size $L=128$, the parameter
$b=3$, while $\rho=4$. The dashed line is a
guide to the eye for the part of $P_{s}(\tau)$ behaving as
$\tau^{-3/2}$. (b) Scaling of the bump in $P_{s}(\tau)$. The data are
obtained from MC simulations.}
\label{ptau}
\end{center}
\end{figure}

Figure \ref{ptau}(a) shows the Monte Carlo results for the survival probability
distribution $P_{s}(\tau)$ of the largest mass, obtained by computing
the distribution of the time interval between successive relocations.
$P_{s}(\tau)$ has two parts, (i) a
power law part $\sim \tau^{-3/2}$ and (ii) another part, which corresponds to
the bump in Fig. \ref{ptau}(a), and has the scaling form
$(\rho-\rho_{c})^{-(b+2)}L^{-(b+1)}f\left(\tau/T_{s}\right)$, as shown
in Fig. \ref{ptau}(b). The prefactor comes from the
normalization of $P_{s}(\tau)$ to unity, with the cutoff for the
$\tau^{-3/2}$ part taken to scale as $T_{r}$. The power law part holds for times when the two
sites with mass $\sim M_{0}/2$ compete to hold the largest mass. The random walk argument
of the preceding paragraph predicts a $\tau^{-3/2}$ decay, since $P_{s}(\tau)$
then stands for the probability for the random walker to cross the origin for the first
time.  The second part in $P_{s}(\tau)$
arises from the relatively long time for which the condensate is
stationary on one site. 

We now explain the behavior of $C(t)/t$ in the different regimes seen in Fig. \ref{ctcond}(a). The condensate is stationary on
one site for a long time $\tau_{1}$, which is a random interval distributed as $p(\tau_{1})
\sim (\rho-\rho_{c})^{-(b+2)}L^{-(b+1)}f\left(\tau_{1}/T_{s}\right)$,
with the characteristic survival time $T_{s}$.
In regime (i), when  $t \ll T_{s}$, the condensate is stationary and acts as a
reservoir for fluctuations, preventing their transport around the system
as a kinematic wave. As a result, Burke's theorem is valid over this
time scale and the
integrated current is Poisson distributed with intensity
$v_{\mathrm{max}}=1$,
implying $C(t)/t=1$. When $t \sim T_{s}$, the
condensate starts to move from one site to another by transferring its mass across the intervening bonds. As a result, these bonds pick up enhanced
fluctuations [$\propto (\Delta M_{0})^{2}$] in the integrated current over the
relocation time interval $\tau_{2}$, which is a random variable with the
characteristic time $T_{r}$. These enhanced fluctuations lead to the
rise in $C(t)/t$ as a function of $t$ in regime (ii). The collapse of the rise
times seen in the scaling plot of Fig. \ref{ctcond}(b) confirms this picture. In regime (iii),
after $t \sim T_{s}+T_{r}$, the condensate has completed relocating,
so current fluctuations revert to Burke-like behavior,
resulting in the fall of $C(t)/t$ in time. The slow fall in regime (iii)
arises from the wide
distribution of the time $\tau_{2}$, and further relocations. To
predict the behavior in regime (iv), where $t \gg
T_{s}+T_{r}$, we construct below a simple relocation model (RM) that
describes the effect of condensate relocation on the long-time
behavior of current fluctuations.

\begin{figure}
\begin{center}
\includegraphics[scale=0.35]{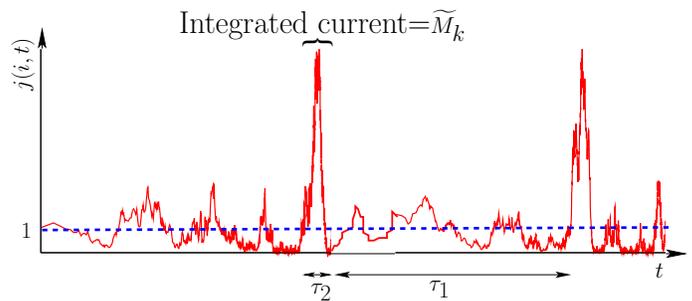}
\caption{(Color online) Schematic plot of the instantaneous
current $j(i,t)$ across the bond $(i, i+1)$ at time $t$ as a
function of time. The random variable $\tau_{1}$ is
the time for which the condensate is stationary on a site, while the
random variable $\tau_{2}$ stands for the relocation time of the
condensate. $\widetilde{M}_{k}$ is the integrated current over time
$\tau_{2}$, arising from the $k$th relocation of the condensate across
the bond $(i, i+1)$.}
\label{rm}
\end{center}
\end{figure}

Figure \ref{rm} shows schematically the instantaneous
current $j(i,t)$ across the bond $(i,i+1)$ as a function of time. In a
given history, let $K$ be the number of condensate relocations in a fixed time $t$. Here, $K$ is a random variable with mean given
approximately by $\langle K \rangle
\approx t/(DT_{s}+BT_{r})$, where $B$ and $D$ are constants, independent of the density and the
system size. At the $k$th relocation of the condensate
($k=1,2,\ldots,K$), let
$\widetilde{M}_{k}$ denote the amount of mass transferred across the bond
$(i,i+1)$ over the interval $\tau_{2}$. The random variable
$\widetilde{M}_{k}$ has an average
$M_{0}$ and variance $\Delta M_{0}$. The integrated current
$H(i,t)$ is given approximately as 
$H(i,t) \approx \sum_{k=1}^{K} ~\widetilde{M}_{k}+\int_{0}^{t-KBT_{r}}dt' j(i,t')$. 
For $b>3$, on computing the variance of the integrated current, we
get $C(t) \approx GL\langle K \rangle +(t-\langle K \rangle B T_{r})$,
where $G$ is a constant. On substituting for
$\langle K \rangle$ and neglecting the time scale $T_{r} \sim
(\rho-\rho_{c})^{2}L^{2}$ in comparison
to $T_{s} \sim (\rho-\rho_{c})^{b+1}L^{b}$, we obtain the asymptotic
behavior
\begin{equation}
C(t) \sim [L^{-\theta}(\rho-\rho_{c})^{-(b+1)}+1]t,  
\label{cktapprox}
\end{equation}
with $\theta=b-1$ for $b>3$. For $b$ between $2$ and $3$, we find that
$\theta=(b^{2}-b-2)/(b-1)$. Thus, the RM predicts that, for all values of $b$, at long
times $t \gg T_{s}+T_{r}$ [regime (iv)], $C(t)/t$ approaches a size-dependent
constant which scales down with the system size. This long-time regime
(iv) could not
be accessed in simulations for the system sizes shown in Fig. \ref{ctcond}(b), but we 
confirmed its existence for smaller systems. 

In summary, we addressed the dynamics of steady state fluctuations of a zero-range process which undergoes a nonequilibrium phase
transition from a disordered to a condensed phase. Different
dynamical properties emerge in the two phases. In the disordered
phase, fluctuations move around the system as a kinematic
wave. Such a wave, though present in the bulk, cannot circulate around
in the condensed phase because the condensate subsumes fluctuations. The dynamics is governed by the condensate relocation through a slower process of transfer of particles
from site to site, contributing enhanced fluctuations to the
particle current across the intervening bonds.

M.B. and S.N.M. acknowledge the Indo-French Centre for
the Promotion of Advanced Research (IFCPAR) for support under Project No. 3404-2.

\end{document}